\documentclass[11pt]{article}

\RequirePackage[OT1]{fontenc}
\RequirePackage[amsthm,amsmath,natbib]{imsart}
\usepackage{amssymb}
\usepackage{graphicx}
\usepackage[latin1]{inputenc}
\usepackage{multirow}

 \newcommand{\eps}{\varepsilon}
 \newcommand{\To}{\longrightarrow}
 \newcommand{\B}{\mathcal{B}}

 \newcommand{\s}{\mathcal{S}}

 \newcommand{\E}{\mathbb{E}}
 
 \newcommand{\X}{\mathcal{X}}
 \newcommand{\F}{\mathcal{F}}

 \newcommand{\Real}{\mathbb{R}}

 \newcommand{\abs}[1]{\left\vert#1\right\vert}

 \newcommand{\norm}[1]{\left\Vert#1\right\Vert}

\newtheorem{theorem}{Theorem}[section]

\newtheorem{algo}{Algorithm}[section]

\theoremstyle{definition}

\newtheorem{remark}{Remark}[section]

\newcounter{hypA}

\newcounter{saveeqn}

\newlength{\defaultheadheight}
\newlength{\defaultheadsep}
\begin{document}

\begin{frontmatter}
\title{Bayesian computation for statistical models with intractable normalizing constants}
\runtitle{Bayesian computation for intractable normalizing constants}

\vspace{1cm}

{\small {\em Yves F.~Atchad\'e\footnote{Department of Statistics, University of Michigan, email:
yvesa@umich.edu},\; Nicolas Lartillot \footnote{LIRM, Universit\'e de Montpellier 2, email: nicolas.lartillot@lirmm.fr} 
and Christian P.~Robert \footnote{CEREMADE, Universit\'e Paris-Dauphine and CREST, INSEE, email: xian@ceremade.dauphine.fr}}\\[5pt]
(April 2008)} \\[10pt]

\begin{abstract}
This paper deals with some computational aspects in the Bayesian analysis of statistical models with intractable normalizing constants. In the presence of intractable normalizing constants in the likelihood function, traditional MCMC methods cannot be applied. We propose an approach to sample from such posterior distributions.  The method can be thought as a Bayesian version of the MCMC-MLE approach of \cite{geyeretthompson92}.
To the best of our knowledge, this is the first general and asymptotically consistent Monte Carlo method for such problems. We illustrate the method with examples from image segmentation and social network modeling. We study as well the asymptotic behavior of the algorithm and obtain a strong law of large numbers for empirical averages.
\end{abstract}



\end{frontmatter}

\section{Introduction}\label{intro}Statistical inference for models with intractable normalizing constants poses a major computational challenge. This problem occurs in the statistical modeling of many scientific problems. Examples include the analysis of spatial point processes (\cite{molleretwaag03}), image analysis (\cite{ibanezetsimo03}), protein design (\cite{lartillotetal06}) and many others.  The problem can be described as follows. Suppose we have a dataset $x_0\in(\X,\B)$ generated from a statistical model $e^{E(x,\theta)}\lambda(dx)/Z(\theta)$ with parameter $\theta\in (\Theta,\Xi)$, where the normalizing constant $Z(\theta)=\int_{\X}e^{E(x,\theta)}\lambda(dx)$ depends on $\theta$ and is not available in closed form. Let $\mu$ be the prior density of the parameter $\theta\in (\Theta,\Xi)$.  The posterior distribution of $\theta$ given $x_0$ is then given by
\begin{equation}\label{pd}\pi(\theta)\propto \frac{1}{Z(\theta)}e^{E(x_0,\theta)}\mu(\theta).\end{equation}
When $Z(\theta)$ cannot be easily evaluated, Monte Carlo simulation from this posterior distribution is problematic even using Markov Chain Monte Carlo (MCMC). \cite{murrayetal06} uses the term \textit{doubly intractable distribution} to refer to posterior distributions of the form (\ref{pd}).
Current Monte Carlo sampling methods do not allow one to deal with such models in a Bayesian framework.
For example, a Metropolis-Hastings algorithm with proposal kernel $Q$ and target distribution $\pi$, would have acceptance ratio $\min\left(1,\frac{e^{E(x_0,\theta')}}{e^{E(x_0,\theta)}}\frac{Z(\theta)}{Z(\theta')}\frac{Q(\theta',\theta)}{Q(\theta,\theta')}\right)$ which cannot be computed as it involves the intractable normalizing constant $Z$ evaluated at $\theta$ and $\theta'$.

An early attempt to deal with this problem is the pseudo-likelihood approximation of Besag (\cite{besag74}) which approximates the model $e^{E(x,\theta)}$  by a more tractable model. Pseudo-likelihood inference provides a first approximation but typically performs poorly (see e.g. \cite{marinetal07}). Maximum likelihood inference is possible. MCMC-MLE, a maximum likelihood approach using MCMC has been developed in the 90's (\cite{geyeretthompson92, geyer94}). Another related approach to find MLE estimates is Younes' 
algorithm (\cite{younes88}) based on stochastic approximation. An interesting  simulation study comparing these three methods is presented in \cite{ibanezetsimo03}.

Comparatively little work has been done to develop asymptotically exact methods for the Bayesian approach to this problem. 
But various approximate algorithms exist in the literature, often based on path sampling (\cite{gelmanetmeng98}).
Recently, \cite{molleretal06} have shown that if exact sampling of $X$ from $e^{E(x,\theta)}/Z(\theta)$ (as a density in $(\X,\B)$) is possible then a valid MCMC algorithm to sample from (\ref{pd}) can be developed. See also \cite{murrayetal06} for some improvements. Their approach uses a clever auxiliary variable algorithm. But intractable normalizing constants often occur in models for which exact sampling of $X$ is not possible or is very expensive. Another recent development to the problem is the approximate Bayesian computation schemes of Plagnol-Tavar\'e (\cite{plagnolettavare04}) but which sample only approximately from the posterior distribution.

In this paper, we propose an adaptive Monte Carlo approach to sample from (\ref{pd}). Our algorithm generates a stochastic process (not Markov in general) $\{(X_n,\theta_n),\;n\geq 0\}$ in $\X\times\Theta$ such that as $n\to\infty$, the marginal distribution of $\theta_n$ converges to (\ref{pd}). It is clear that any method to sample from (\ref{pd}) will have to deal with the intractable normalizing constant $Z(\theta)$. In the auxiliary variable method of \cite{molleretal06}, computing $Z(\theta)$ is replaced in a sense by perfect sampling from $e^{E(x,\theta)}/Z(\theta)$. This strategy works well so long as perfect sampling is feasible and inexpensive.
In the present work, we take another approach building on the idea of estimating the entire function $Z$ from a single Monte Carlo sampler. The starting point of the method is importance sampling. Suppose that for some $\theta^{(0)}\in\Theta$, we can sample (perhaps by MCMC) from the density $e^{E(x,\theta^{(0)})}/Z(\theta^{(0)})$ in $(\X,\B)$. Using this sample, we can certainly estimate $Z(\theta)/Z(\theta^{(0)})$ for any $\theta\in\Theta$. This is the same idea behind the MCMC-MLE algorithm of \cite{geyeretthompson92}. But it is well known that these estimates are typically very poor as $\theta$ gets far from $\theta^{(0)}$. Now, suppose that instead of a single point $\theta^{(0)}$, we generate a population $\{\theta^{(i)},\; i=1,\ldots,d\}$ in $\Theta$ and that we can sample from $\Lambda^*(x,i)\propto e^{E(x,\theta^{(i)})}/Z(\theta^{(i)})$ on $\X\times\{1,\ldots,d\}$. Then we show that in principle, efficient estimation for $Z(\theta)$ is possible for any $\theta\in\Theta$. 
Building on \cite{atchadeetliu04} and the ideas sketched above, we propose an algorithm that generates a random process $\{(X_n,\theta_n),\;n\geq 0\}$ such that the marginal distribution of $X_n$ converges to $\Lambda^*$ and  the marginal distribution of $\theta_n$ converges to (\ref{pd}). This random process is not a Markov chain in general but we show (from first principle) that $\{\theta_n\}$ has limiting distribution $\pi$ and satisfies a strong law of large numbers.

The paper is organized as follows. A full description of the method including practical implementation details is given in Section \ref{wl}. We illustrate the algorithm with three examples. The Ising model, a Bayesian image segmentation example and a Bayesian modeling of social networks. The examples are presented in Section \ref{examples}. Some theoretical aspects of the method are discussed in Section \ref{theory} with the proofs postponed to \ref{proofs}.

\section{Sampling from posterior distributions with intractable normalizing constants}\label{wl}
Throughout, we fix the sample space $(\X,\B,\lambda)$ and the parameter space $(\Theta,\Xi)$. The problem of interest is to sample from the posterior distribution (\ref{pd}) with 
\begin{equation}\label{Zdecomp0}Z(\theta)=\int_{\X}e^{E(x,\theta)}\lambda(dx).\end{equation}
Let $\{\theta^{(i)},\;i=1,\ldots,d\}$ be a sequence of $d$ points in $\Theta$. Let $\Lambda^*$ be the probability measure on $\X\times\{1,\ldots,d\}$ given by:
\begin{equation}\label{pistar}\Lambda^*(x,i)=\frac{e^{E(x,\theta^{(i)})}}{dZ(\theta^{(i)})},\;\;x\in\X,i\in\{1,\ldots,d\}.\end{equation}
 Let $\kappa(\theta,\theta')$ be a similarity kernel on $\Theta\times\Theta$ such that $\sum_{i=1}^d\kappa(\theta,\theta^{(i)})=1$ for all $\theta\in\Theta$. The starting point of the algorithm is the following decomposition of the partition function:
\begin{eqnarray}\label{Zdecomp}Z(\theta)&=&\int_{\X}e^{E(x,\theta)}\lambda(dx)\nonumber\\
&=&\sum_{i=1}^d\kappa(\theta,\theta^{(i)})\int_{\X}e^{E(x,\theta)-E(x,\theta^{(i)})}e^{E(x,\theta^{(i)})}\lambda(dx)\nonumber\\
&=&d\sum_{i=1}^d\kappa(\theta,\theta^{(i)})Z(\theta^{(i)})\int_{\X}e^{E(x,\theta)-E(x,\theta^{(i)})}\frac{e^{E(x,\theta^{(i)})}}{dZ(\theta^{(i)})}\lambda(dx)\nonumber\\
&=&\sum_{i=1}^d\int_{\X}\Lambda^*(x,i)h_\theta(x,i)\lambda(dx),\end{eqnarray}
where 
\begin{equation}\label{htheta}h_\theta(x,i)=d\kappa(\theta,\theta^{(i)})Z(\theta^{(i)})e^{E(x,\theta)-E(x,\theta^{(i)})}.\end{equation}

The interest of the decomposition (\ref{Zdecomp}) is that $\{Z(\theta^{(i)})\}$ and $\Lambda^*$ do not depend on $\theta$. Therefore, using samples from $\Lambda^*$, this decomposition gives an approach to estimate $Z(\theta)$ for all $\theta\in\Theta$. This estimate should be reliable provided $\theta$ is close to at least one particle $\theta^{(i)}$. The problem of sampling from probability measures such as $\Lambda^*$ has been recently considered by \cite{atchadeetliu04} building on the Wang-Landau algorithm of \cite{wangetlandau01}. We follow and improve that approach here. The resulting estimate of $Z(\theta)$ can then continuously be fed to a second Monte Carlo sampler that carries the simulation  with respect to $\pi$. This suggests an adaptive Monte Carlo sampler to sample from (\ref{pd}) which we develop next.

For any $c=(c(1),\ldots,c(d))\in\Real^d$, we define the following density function on $\X\times\{1,\ldots,d\}$: 
\begin{equation}\Lambda_c(x,i)\propto e^{E(x,\theta^{(i)})-c(i)}.\end{equation} 
With  $c=\log(Z)$, $\Lambda_c=\Lambda^*$. The reader should think of $c$ as an estimate of $z$, with $z(i):=\log Z(\theta^{(i)})$. The algorithm will adaptively adjust $c$ such that the marginal distribution on $\{1,\ldots,d\}$ is approximately uniform. In which case, we should have $c(i)=\log Z(i)$. Let $\{\gamma_n\}$ be a sequence of (possibly random) positive numbers. 
We propose a non-Markovian adaptive sampler that lives in $\X\times\{1,\ldots,d\}\times\Real^{d}\times\Theta$. We start from an initial state $(X_0,I_0,c_0,\theta_0)\in\X\times\{1,\ldots,d\}\times\Real^d\times\Theta$, where $c_0\in\Real^d$ is the initial estimate of $z$. For example, $c_0\equiv 0$. At time $n$, given $(X_n,I_n,c_n,\theta_n)$ we first generate $X_{n+1}$ from $P_{I_n}(X_n,\cdot)$, where $P_{i}$ is a transition kernel on $(\X,\B)$ with invariant distribution $e^{E(x,\theta^{(i)})}/Z(\theta^{(i)})$. Next, we generate $I_{n+1}$ from the distribution on $\{1,\ldots,d\}$ proportional to $e^{E(X_{n+1},\theta^{(i)})-c_n(i)}$. Then we update the current estimate of $\log(Z)$ to $c_{n+1}$ given by:
\begin{equation}\label{recurc}c_{n+1}(i)=c_n(i)+\gamma_n \frac{e^{E(X_{n+1},\theta^{(i)})-c_n(i)}}{\sum_{j=1}^de^{E(X_{n+1},\theta^{(j)})-c_n(j)}},\;i=1,\ldots,d.\end{equation}
In view of (\ref{Zdecomp}), we can estimate  $Z(\theta)$ by:
\begin{equation}\label{Zn}Z_{n+1}(\theta)=\sum_{i=1}^d\kappa(\theta,\theta^{(i)})e^{c_{n+1}(i)}\left[\frac{\sum_{k=1}^{n+1}e^{E(X_k,\theta)-E(X_k,\theta^{(i)})}\textbf{1}_{i}(I_k)}{\sum_{k=1}^{n+1}\textbf{1}_{i}(I_k)}\right],\end{equation} with the convention that $0/0=0$. Finally, for any positive function $\zeta:\Theta\to (0,\infty)$, let $Q_\zeta$ be a transition kernel on $(\Theta,\Xi)$ with invariant distribution 
\begin{equation}\label{pdn}\pi_{\zeta}(\theta)\propto \frac{1}{\zeta(\theta)}e^{E(x_0,\theta)}\mu(\theta).\end{equation}
Given $(X_{n+1},I_{n+1},c_{n+1},Z_{n+1},\theta_n)$, we generate $\theta_{n+1}$ from $Q_{Z_{n+1}}(\theta_n,\cdot)$, where $Z_{n+1}$ is the function defined by (\ref{Zn}).

The algorithm can be summarized as follows.
\begin{algo}\label{awl}. Let $(X_0,I_0,c_0,\theta_0)\in\X\times\{1,\ldots,d\}\times\Real^d\times\Theta$ be the initial state of the algorithm. Let $\{\gamma_n\}$ be (a possibly random) sequence of positive numbers. At time $n$, given $(X_n,I_n,c_n,\theta_n)$:
\begin{description}
\item [1.] Generate $X_{n+1}$ from $P_{I_n}(X_n,\cdot)$ where $P_{i}$ is any ergodic kernel on $(\X,\B)$ with invariant distribution $e^{E(x,\theta^{(i)})}/Z(i)$.
\item [2.] Generate $I_{n+1}$ by sampling from the distribution on $\{1,\ldots,d\}$ proportional to $e^{E(X_{n+1},\theta^{(i)})-c_n(i)}$.
\item [3.] Compute $c_{n+1}$, the new estimate of $g$ using (\ref{recurc}).
\item [4.] Using the function $Z_{n+1}$ defined by (\ref{Zn}), generate $\theta_{n+1}$ from $Q_{Z_{n+1}}(\theta_n,\cdot)$.
\end{description}
\end{algo}

\begin{remark}
\begin{enumerate}
\item The algorithm can be seen as an MCMC-MCMC analog to the MCMC-MLE of \cite{geyeretthompson92}. Indeed, with $d=1$, the decomposition (\ref{Zdecomp}) becomes 
\[Z(\theta)/Z(\theta^{(1)})=\E\left[e^{E(\theta,X)-E(X,\theta^{(1)})}\right],\]
where the expectation is taken with respect to the density $e^{E(x,\theta^{(1)})}/Z(\theta^{(1)})$. But as discussed in the introduction, when $E(\theta,X)-E(X,\theta^{(1)})$ has a large variance, the resulting estimate is terribly poor. 
\item We introduce $\kappa$ to serve as a smoothing factor so that the particles $\theta^{(i)}$'s close to $\theta$ contribute more to the estimation of $Z(\theta)$. We expect this  smoothing step to reduce the variance of the overall estimate of $Z(\theta)$. In the simulations we choose 
\[\kappa(\theta,\theta^{(i)})=\frac{e^{-\frac{1}{2h^2}\norm{\theta-\theta^{(i)}}^2}}{\sum_{j=1}^de^{-\frac{1}{2h^2}\norm{\theta-\theta^{(j)}}^2}}.\]
The value of the smoothing parameter $h$ is set by trials and errors for each example. 
\item The implementation of the algorithm requires keeping track of all the samples $X_k$ that are
generated (Equation (\ref{Zn})). $\X$ can be a very high-dimensional space and we are aware of the fact that in practice, this bookkeeping can significantly slow down the algorithm. But in many cases, the function $E$ takes the form $E(x,\theta)=\sum_{l=1}^KS_l(x)\theta_l$ for some real-valued functions $S_l$. In these cases, we only need to keep track of the statistics $\{\left(S_1(X_n),\ldots,S_K(X_n)\right),\;n\geq 0\}$. All the examples discussed in the paper fall in this latter category.
\item As mentioned earlier, the update of $(X_n,I_n,c_n)$ is essentially the Wang-Landau algorithm of \cite{atchadeetliu04} with the following important difference. \cite{atchadeetliu04} propose to update $c_n$ one component per iteration:
\[c_{n+1}(i)=c_n(i)+\gamma_n\textbf{1}_{\{i\}}(I_{n+1}).\]
We improve on this scheme in (\ref{recurc}) by Rao-Blackwellization  where we integrate out $I_{n+1}$. 
\item As mentioned above, and we stress this again, this algorithm is not Markovian in any way. The process $\{(X_n,I_n,c_n)\}$ is not a Markov chain but a nonhomogeneous Markov chain if we let $\{\gamma_n\}$ be a deterministic sequence. $\{\theta_n\}$, the main random process of interest is not a Markov chain either. Nevertheless, the marginal distribution of $\theta_n$ will typically converge to $\pi$. This is because, $Q_{Z_n}$, the conditional distribution of $\theta_{n+1}$ given $\F_{n}$ converges to $Q_Z$ as $n\to\infty$ and $Q_Z$ is a kernel with invariant distribution $\pi$. We make this precise by showing that a strong law of large numbers holds for additive functionals of $\{\theta_n\}$.
\end{enumerate}
\end{remark}
We now discuss the choice of the parameters of the algorithm.
\subsection{Choosing $d$ and the particles $\{\theta^{(i)}\}$}
We do not have any general approach in choosing $d$ and $\{\theta^{(i)}\}$ but we give some guidelines. The general idea is that the particles $\{\theta^{(i)}\}$ need to cover reasonably well the range of the density $\pi$ and be such that for any $\theta\in\Theta$, the density $e^{E(x,\theta)}/Z(\theta)$ in $\X$ can be well approximated by at least one of the densities $e^{E(x,\theta^{(i)})}/Z(i)$. One possibility is to sample $\theta^{(i)}$ independently from the prior distribution $\mu$, some tempered version of it or some other similar distribution. We follow this approach in the examples below. Another possibility is to use a grid of points in $\Theta$. The value of $d$, the number of particles, should depend on the size of $\Theta$. We need to choose $d$ such that the distributions $e^{E(x,\theta^{(i)})}/Z(i)$ (in $\X$) overlap. Otherwise, estimating the constants $\{Z(i)\}$ can be difficult. This implies that  $d$ should not be too small. In the simulation examples be!
 low we choose $d$ between $100$ and $500$.

\subsection{Choosing the step-size $\{\gamma_n\}$}\label{gam}
It is shown in \cite{atchadeetliu04} that the recursion (\ref{recurc}) can also be written as a stochastic approximation algorithm with step-size $\{\gamma_n\}$, so that in theory, any positive sequence $\{\gamma_n\}$ such that $\sum\gamma_n=\infty$ and $\sum\gamma_n^2<\infty$ can be used. But the convergence of $c_n$ to $\log Z$ is very sensitive to the choice $\{\gamma_n\}$. If the $\gamma_n$'s are overly small, the recursive equation in (\ref{recurc}) will make very small steps. But if these numbers are overly large, the algorithm will have a large variance. In both cases, the convergence to the solution will be slow. Overall, choosing the right step-size for a stochastic approximation algorithm is a difficult problem. Here we follow \cite{atchadeetliu04} which has elaborated on a  heuristic approach to this problem originally proposed by \cite{wangetlandau01}. 

The main idea of the method is that typically, the larger $\gamma_n$, the easier it is for the algorithm to move around the state space. Therefore, at the beginning $\gamma_0$ is set at a relatively large value. This value is kept until $\{I_n\}$ has visited equally well all the points of $\{1,\ldots,d\}$.  Let $\tau_1$ be the first time where the occupation measure of $\{1,\ldots,d\}$ by $\{I_n\}$ is approximately uniform. Then we set $\gamma_{\tau_1+1}$ to some smaller value (for example $\gamma_{\tau_1+1}=\gamma_{\tau_1}/2$) and the process is iterated until $\gamma_n$ become sufficiently small. As which point we can choose to switch to a deterministic sequence of the form $\gamma_n=n^{-1/2-\eps}$. Combining this idea with Algorithm \ref{awl}, we get the following.

\begin{algo}\label{awl2}. Let $\gamma>\eps_1>0$, $\eps_2>0$ be constants and let $(X_0,I_0,c_0,\theta_0)$ be some arbitrary initial state of the algorithm. Set $v=0\in\Real^d$ and $n=0$. While $\gamma>\eps_1$ and given $\F_n=\sigma\{(X_k,I_k,c_k,\theta_k),\;k\leq n\}$, 
\begin{description}
\item [1.] Generate $(X_{n+1},I_{n+1},c_{n+1},\theta_{n+1})$ as in Algorithm \ref{awl}.
\item [2.] For $i=1,\ldots,d$: set $v(i)=v(i)+\textbf{1}_{i}(I_{n+1})$.
\item [3.] If $\max_{i}\abs{v(i)-\frac{1}{d}}\leq \frac{\eps_2}{d}$, then set $\gamma=\gamma/2$ and set $v=0\in\Real^d$.
\end{description}
\end{algo}

\begin{remark} We use this algorithm in the examples below with the following specifications. We set the initial $\gamma$ to $1$, $\eps_2=0.2$. We run $\{(X_n,I_n,c_n)\}$ until $\gamma\leq \eps_1=0.001$ before starting $\{\theta_n\}$ and switching to a deterministic sequence $\gamma_n=\eps_1/n^{0.7}$.
\end{remark}

%
\section{Convergence analysis}\label{theory}
In this section, we derive a law of large numbers under some verifiable conditions. The process of interest here is $\{\theta_n\}$. Let $(\Omega,\F,\Pr)$ be the reference probability triplet. We equip $(\Omega,\F,\Pr)$ with the filtration $\{\F_n\}$, where $\F_n=\sigma\{(X_{k+1},I_{k+1},c_{k+1},\theta_{k}),\;k\leq n\}$. Note that $\F_n$ includes $(X_{n+1},I_{n+1},c_{n+1})$ since these random variables are used in generating $\theta_{n+1}$. From the definition of the algorithm, we have:
\begin{equation}\label{dyn}\Pr\left(\theta_{n+1}\in A\vert \F_n\right)=Q_{Z_{n+1}}(\theta_n, A),\;\;\Pr-a.s..\end{equation}
We see from (\ref{dyn}) that $\{\theta_n,\F_n\}$ is an adaptive Monte Carlo algorithm with varying target distribution. 
 In analyzing $\{\theta_n,\F_n\}$ here, we do not strive for the most general result but restrict ourself to conditions that can be easily checked in the examples considered in the paper. We assume that $\Theta$ is a compact subset of $\Real^q$, the $q$-dimensional Euclidean space equipped with its Borel $\sigma$-algebra and the Lebesgue measure.  Firstly, we assume that the function $E$ is bounded:

\noindent\textbf{(A1)}: \textit{There exist $m,M\in\Real$ such that:
\begin{equation}\label{a1}m\leq E(x,\theta)\leq M,\;\;x\in\X,\theta\in\Theta.\end{equation}}

In many applications, and this is the case for the examples discussed below, $\X$ is a finite set (typically very large) and $\Theta$ is a compact set. In these cases, (A1) is easily checked. In order to proceed any further, we need some notations. A transition kernel on $(\X,\B)$ operates on measurable real-valued functions $f$ as $Pf(x)=\int P(x,dy)f(y)$, and the product of two transition kernels $P_1$ and $P_2$ is the transition kernel defined as $P_1P_2(x,A)=\int P_1(x,dy)P_2(y,A)$. We can then define recursively $P^n=PP^{n-1}$, $n\geq 1$, $P^0(x,A)=\textbf{1}_A(x)$. For two probability measures $\mu,\nu$, the total variation distance between $\mu$ and $\nu$ is defined as $\norm{\mu-\nu}_{TV}:=\sup_{A}\abs{\mu(A)-\nu(A)}$. We say that a transition kernel $P$ is geometrically ergodic if $P$ is $\phi$-irreducible, aperiodic and has an invariant distribution $\pi$ such that:
\[\norm{P^n(x,\cdot)-\pi}_{TV}\leq M(x)\rho^n,\;\;n\geq 0\]
for some $\rho\in(0,1)$ and some function $M:\;\X\to(0,\infty]$.

Our next assumption involves the transition kernel $Q_\zeta$.

\noindent\textbf{(A2)}: \textit{For $\zeta:\Theta\to (0,\infty)$, $Q_\zeta$ is a Metropolis kernel with invariant distribution $\pi_\zeta$ and proposal kernel density $p$. There exist $\eps>0$ and an integer $n_0\geq 1$ such that for all $\theta,\theta'\in\Theta$:
\begin{equation}\label{a2}p^{n_0}(\theta,\theta')\geq \eps.\end{equation}}

\begin{remark}\label{rema2} 
\begin{enumerate}
\item The condition (\ref{a2}) clearly holds for most symmetric proposal kernels $p(\theta,\theta')$, provided that
$p(\theta,\theta')$ remains bounded away from $0$ on some ball centered at $\theta$. 
\item (\ref{a2}) often implies that $Q_\zeta$ is uniformly ergodic:
\begin{eqnarray*}Q_\zeta(\theta,A)&\geq& \int_A\min\left(1,e^{E(x_0,\theta')-E(x_0,\theta)}\frac{\zeta(\theta')}{\zeta(\theta)}\right)p(\theta,\theta')d\theta'\\
&\geq& e^{m-M}\inf_{\theta,\theta'\in\Theta}\left(\frac{\zeta(\theta)}{\zeta(\theta')}\right)\int_Ap(\theta,\theta')d\theta'.\end{eqnarray*}
Therefore, provided $\inf_{\theta,\theta'\in\Theta}\left(\frac{\zeta(\theta)}{\zeta(\theta')}\right)>0$, if (\ref{a2}) hold then $Q^{n_0}_\zeta(\theta,A)\geq \eps'\mu_{Leb}(A)$ for some $\eps'>0$.
\end{enumerate}
\end{remark}

\noindent\textbf{(A3)}: \textit{$\{\gamma_n\}$ is a random sequence, adapted to $\{\F_n\}$ such that $\gamma_n>0$, $\sum\gamma_n=\infty$ and $\sum\gamma_n^2<\infty$ $\Pr$-a.s.}

\begin{theorem}\label{thm1}
Assume (A1)-(A3). Assume also that each kernel $P_i$ on $(\X,\B)$ is geometrically ergodic.  Let $h:\;(\Theta,\Sigma)\to\Real$ such that $\abs{h}\leq 1$. Then:
\begin{equation}\frac{1}{n}\sum_{k=1}^nh(\theta_k)\to \pi(h),\;\Pr-a.s.\end{equation}
\end{theorem}
\begin{proof}See Section \ref{proofs}.\end{proof}

\section{Examples}\label{examples}
\subsection{Ising model}\label{ising1d}
We test the algorithm with the Ising model on a rectangular lattice. This is a simulated example. The model is given by $e^{\theta E(x)}/Z(\theta)$
 where 
\begin{equation}\label{E1}E(x)=\sum_{i=1}^m\sum_{j=1}^{n-1}x_{ij}x_{i,j+1}+\sum_{i=1}^{m-1}\sum_{j=1}^{n}x_{ij}x_{i+1,j},\end{equation}
 and $x_i\in\{1,-1\}$. We use $m=n=64$. We generate the data $x_0$ from $e^{\theta E(x)}/Z(\theta)$ with $\theta=0.40$ by perfect sampling using the Propp-Wilson algorithm. Using $x_0$ and postulating the model $e^{\theta E(x)}/Z(\theta)$, we would like to infer on $\theta$. We use the prior $\mu(\theta)=\textbf{1}_{(0,1)}(\theta)$. We set $d=100$ and generate the points $\{\theta^{(i)}\}$ independently and uniformly in $(0,1)$. As described in Section \ref{gam}, we use the flat histogram approach in selecting $\{\gamma_n\}$ with an initial value $\gamma_0=1$, until $\gamma_n$ becomes smaller than $0.001$. Then we start feeding the adaptive chain $\{\theta_n\}$ which is run for $10,000$ iterations. In updating $\theta_n$, we use a Random Walk Metropolis sampler with proposal distribution $\mathcal{U}(\theta_n-b,\theta_n+b)$ (with reflexion at the boundaries) for some $b>0$. We adaptively update $b$ so as to reach an acceptance rate of $30\%$ (see e.g. \cite{atchade05}). We d!
 iscard th!
 e first $1,999$ points as a burn-in period. The results are  plotted on Figure 1. As we can see from these plots, the sampler appears to have converged to the posterior distribution $\pi$. The mixing rate of the algorithm as inferred from the autocorrelation function seems fairly good. In addition, the algorithm yields an estimate of the partition function $\log Z(\theta)$ which can be re-used in other sampling problems.

\begin{center}
\scalebox{0.5}{\includegraphics{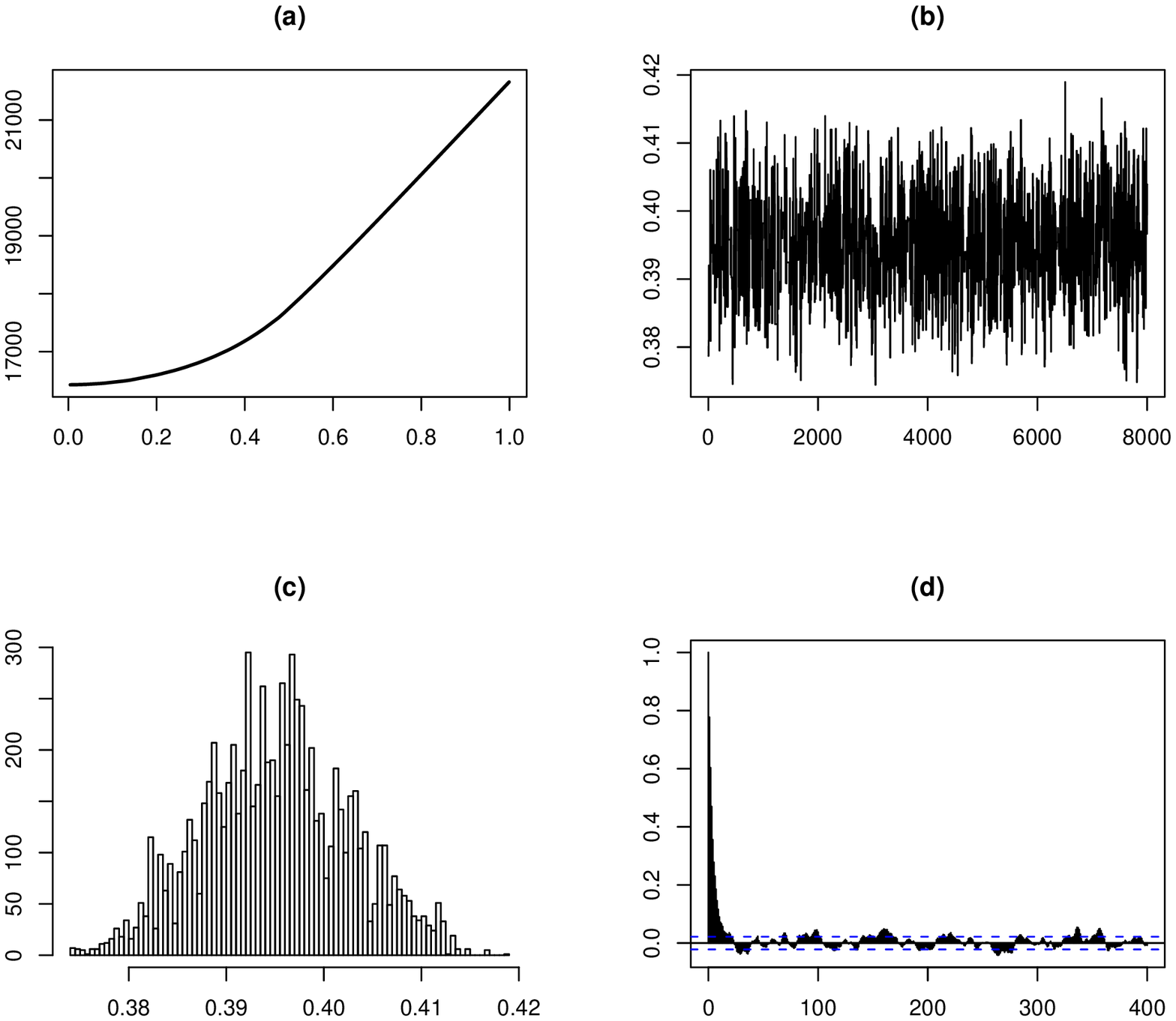}}\\
\noindent\underline{Figure 1}: Output for the Ising model $\theta=0.40$, $m=n=64$. (a): estimation of $\log Z(\theta)$ up to an additive constant; (b)-(d): Trace plot, histogram and autocorrelation function of the adaptive sampler $\{\theta_n\}$.
\end{center} 

\subsection{An application to image segmentation}\label{image}
We use the Ising model above to illustrate an application of the methodology to image segmentation. In image segmentation, the goal is the reconstruction of images from noisy observations (see e.g. \cite{ibanezetsimo03, hurnetal03}). We represent the image by a vector $x=\{x_i,\;i\in\s\}$ where $\s$ is a $m\times n$ lattice and $x_i\in\{1,\ldots,K\}$. Each $i\in\s$ represents a pixel, and $x_i$ is the color of the pixel $i$. $K$ is the number of colors. Here we assume that $K=2$ and $x_i\in\{-1,1\}$ is either black or white. In the image segmentation problem, we do not observe $x$ but a noisy approximation $y$. We assume that:
\begin{equation}\label{condy}y_i\vert x,\sigma^2\stackrel{ind}{\sim}\mathcal{N}(x_i,\sigma^2),\end{equation}
for some unknown parameter $\sigma^2$. Even though (\ref{condy}) is a continuous model, it has been shown to provide a relatively good framework for image segmentation problems with multiple additive sources of noise (\cite{ibanezetsimo03}).  

We assume that the true image $x$ is generated from an Ising model (see Section \ref{ising1d}) with interaction parameter $\theta$. We assume that $\theta$ follows a uniform prior distribution on $(0,1)$ and that $\sigma^2$ has a prior distribution that is proportional to $1/\sigma^2\textbf{1}_{(0,\infty)}(\sigma^2)$. The posterior distribution $(\theta,\sigma^2,x)$ is then given by:
\begin{equation}\label{postimage}\pi\left(\theta,\sigma^2,x\vert y\right)\propto\left(\frac{1}{\sigma^2}\right)^{\frac{\abs{\s}}{2}+1}\frac{e^{\theta E(x)}}{Z(\theta)}e^{-\frac{1}{2\sigma^2}\sum_{s\in\s}\left(y(s)-x(s)\right)^2}\textbf{1}_{(0,1)}(\theta)\textbf{1}_{(0,\infty)}(\sigma^2),\end{equation}
where $E$ is as in (\ref{E1}).

We sample from this posterior distribution using the adaptive chain $\{(y_n,i_n,c_n,\theta_n,\sigma_n^2,x_n)\}$. The chain $\{(y_n,i_n,c_n)\}$ is updated following Steps
(1)-(3) of Algorithm \ref{awl}. It is used to form the adaptive estimate of $Z(\theta)$  as given by (\ref{Zn}) (with $\{y_n,i_n\}$ replacing $\{X_n,I_n\}$). These estimates are used to update $(\theta_n,\sigma_n^2,x_n)$ using a Metropolis-within-Gibbs
scheme. More specifically, given $\sigma_n^2,x_n$, we sample $\theta_{n+1}$ with a Random Walk Metropolis with proposal $\mathcal{U}(\theta_n-b,\theta_n+b)$ (with reflexion at the boundaries) and target proportional to $\frac{e^{\theta E(x_n)}}{Z_n(\theta)}$. Given $\theta_{n+1},x_n$, we generate $\sigma^2_{n+1}$ by sampling from the Inverse-Gamma distribution with parameters $(\frac{\abs{\s}}{2},\frac{1}{2}\sum_{s\in\s}\left(y(s)-x(s)\right)^2)$. And given $(\theta_{n+1},\sigma_{n+1})$, we sample each $x_{n+1}(s)$ from its conditional distribution given $\{x(u) ,\;u\neq s\}$. This conditional distribution is given by 
\[p(x\left(s)=a\vert x(u),u\neq s\right)\propto \exp\left(\theta a\sum_{u\sim s}x(u)-\frac{1}{2\sigma^2}(y(s)-a)^2\right),\;\;a\in\{-1,1\}.\]
Here $u\sim v$ means that pixels $u$ and $v$ are neighbors. 

To test this algorithm, we generate a simulated dataset $y$ according to model (\ref{condy}) with $x$ generated from $e^{\theta E(x)}/Z(\theta)$ by perfect sampling. We use $m=n=64$, $\theta=0.40$ and $\sigma=0.5$. For the implementation details of the algorithm, we make exactly the same choices as in Example \ref{ising1d} above. In particular we choose $d=100$ and generate $\{\theta^{(i)}\}$ uniformly in $(0,1)$. The results are given in Figure 2. Once again, the sample path obtained from $\{\theta_n\}$ clearly suggests that the distribution of $\theta_n$ has converged to $\pi$ with a good mixing rate, as inferred from the autocorrelation plots. 

\begin{center}
\scalebox{0.5}{\includegraphics{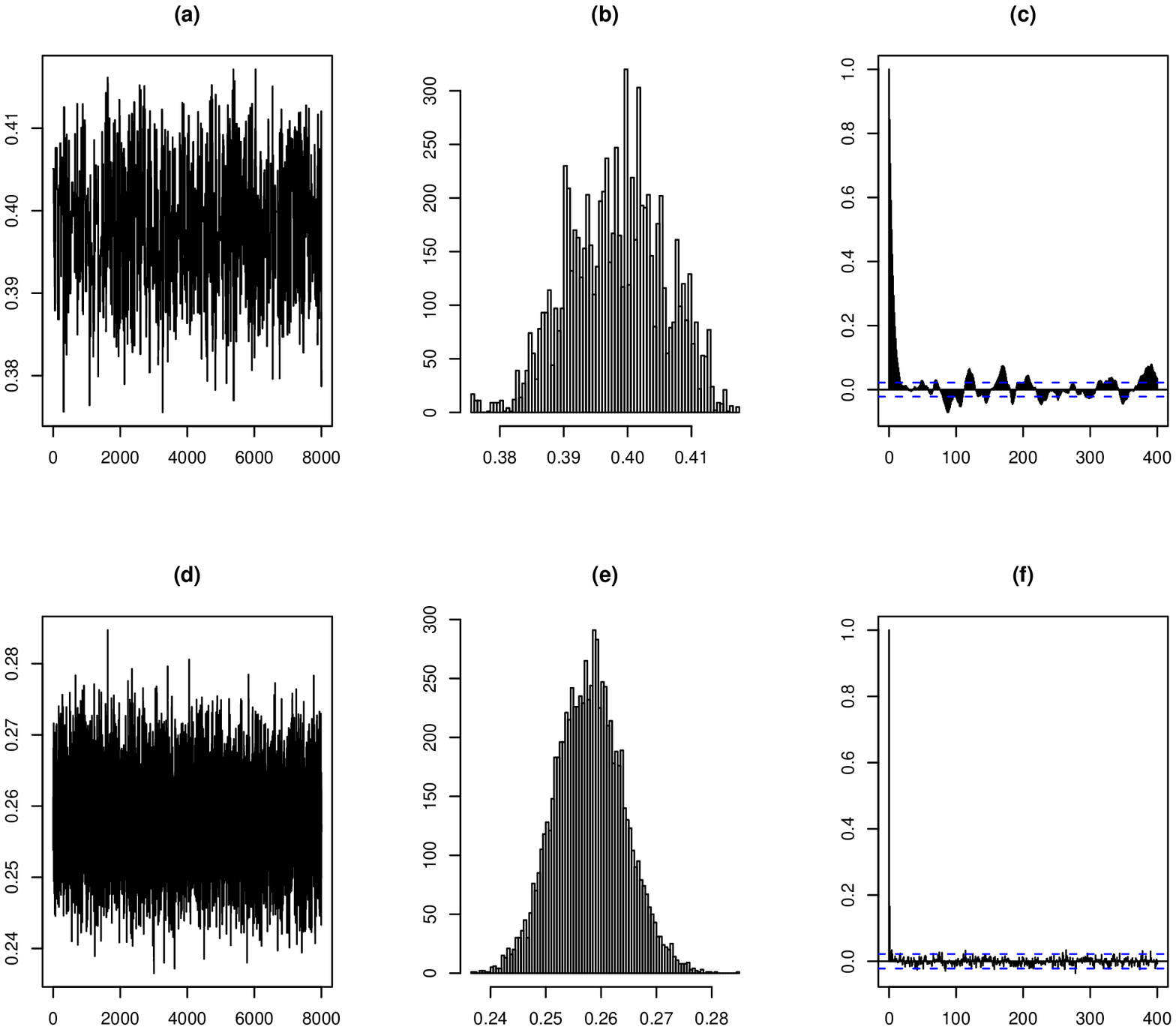}}\\
\noindent\underline{Figure 2}: Output for the image segmentation model. (a)-(c): plots of $\{\theta_n\}$; (d)-(f): plots of $\{\sigma_n\}$.
\end{center} 

\subsection{Social network modeling}\label{socialnet}
We now give an application of the method to a Bayesian analysis of social networks. Statistical modeling of social network is a growing subject in social science (See e.g. \cite{robinsetal07} and the references therein for more details). The set up is the following. We have $n$ actors $I=\{1,\ldots,n\}$. For each pair $(i,j)\in I\times I$, define $y_{ij}=1$ if actor $i$ has ties with actor $j$ and $y_{ij}=0$ otherwise. In the example below, we only consider the case of a symmetric relationship where $y_{ij}=y_{ji}$ for all $i,j$. The ties referred to here can be of various natures. For example, we might be interested in modeling how friendships build up between co-workers or how research collaboration takes place between colleagues. Another interesting example from political science is modeling  co-sponsorship ties (for a given piece of legislation) between members of a house of representatives or parliment. 

In this example we study the Medici business network dataset taken from \cite{robinsetal07} which describes the business ties between 16 Florentine families. Numbering arbitrarily the family from $1$ to $16$, we plot the observed social network in Figure 3. The dataset contains relatively few ties between families and even fewer transitive ties. 

\begin{center}
\scalebox{1}{\includegraphics{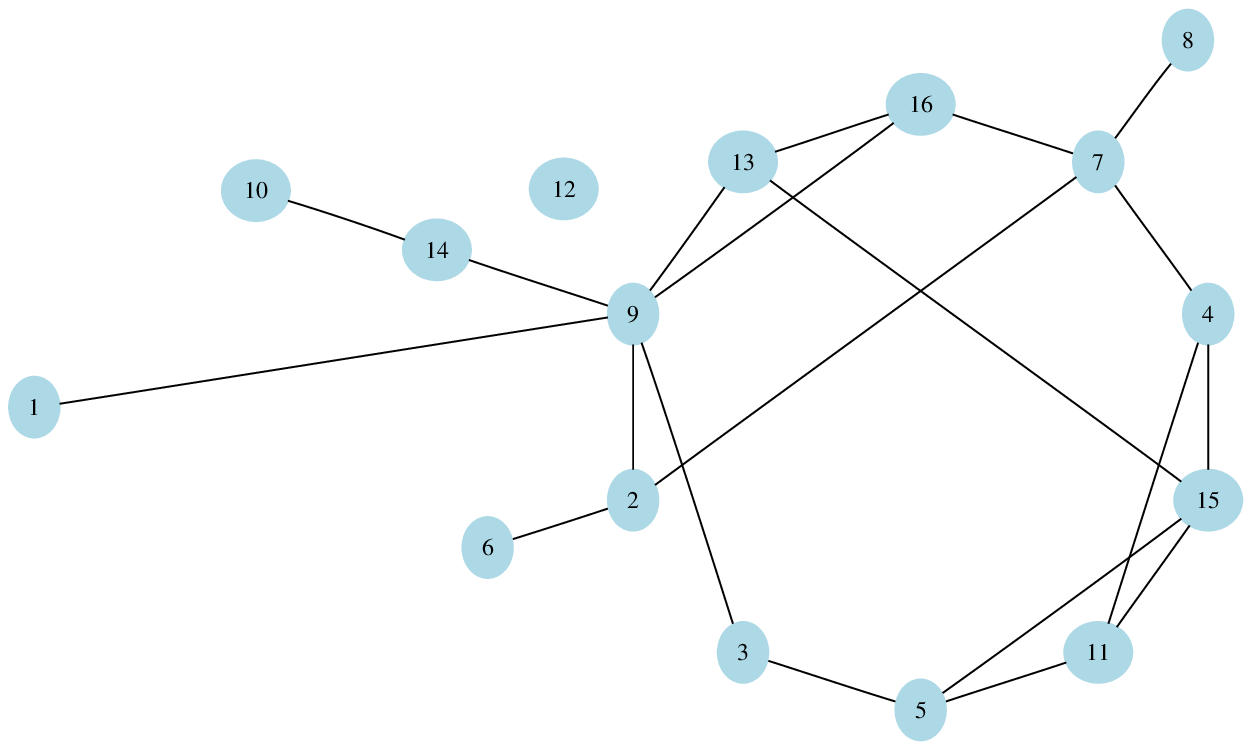}}\\
\noindent\underline{Figure 3}: Business Relationships between $16$ Florentine families. 
\end{center} 

\vspace{0.3cm}
One of the most popular models for social networks is the class of exponential random graph models. In these models, we assume that $\{y_{ij}\}$ is a sample generated from the distribution
\begin{equation}p\left(y\vert \theta_1,\ldots,\theta_K\right)\propto exp\left(\sum_{i=1}^K\theta_iS_i(y)\right),\end{equation}
for some parameters $\theta_1,\ldots,\theta_K$; where $S_i(y)$ is a statistic used to capture some aspect of the network. For this example, and following \cite{robinsetal07}, we consider a $4$-dimensional model with statistics
\[S_1(y)=\sum_{i<j}y_{ij},\;\;\mbox{ the total number of ties},\]
\[S_2(y)=\sum_{i<j<k}y_{ik}y_{jk}
,\;\;\mbox{ the number of two-stars},\]
\[S_3(y)=\sum_{i<j<k<l}y_{il}y_{jl}y_{kl}
,\;\;\mbox{ the number of three-stars},\]
\[S_4(y)=\sum_{i<j<k}y_{ik}y_{jk}y_{ij},\;\;\mbox{ the number of transitive ties}.\]
We assume a uniform prior distribution on $D=(-50,50)^4$  for $\theta=(\theta_1,\theta_2,\theta_3,\theta_4)$ and the posterior distribution writes:

\begin{equation}\label{pisn}\pi\left(\theta\vert y\right)\propto \frac{1}{Z(\theta)}\exp\left(\sum_{k=1}^4\theta_kS_k(y)\right)\textbf{1}_D(\theta).\end{equation}
 We use Algorithm \ref{awl} to sample from (\ref{pisn}).  For this example, we use $400$ particles $\{\theta^{(l)}\}$ generated from a $N(0,5)$ the normal distribution with mean $0$ and variance $5$.  We use the same parametrization as in the previous examples to update $(X_n,I_n,c_n)$. For the adaptive chain $\{\theta_n\}$ we use a slightly different strategy. It turns out that some of the components of the target distribution $\pi$ are strongly related. Therefore we sample from $\pi$ in one block, using a Random Walk Metropolis with a Gaussian kernel $N(0,\sigma^2\Sigma)$ (restricted to $D$) for some $\sigma>0$ and a positive definite matrix $\Sigma$. We adaptively set $\sigma$ so as to reach an acceptance rate of $30\%$. Ideally, we would like to choose $\Sigma=\Sigma_\pi$ the variance-covariance of $\pi$ which of course, is not available. We adaptively estimate $\Sigma_\pi$ during the simulation as in \cite{atchade05}. As before, we run $(X_n,I_n,c_n)$ until $\gamma_n<0.!
 001$. Then we start $\{\theta_n\}$ and run the full chain $(X_n,I_n,c_n,\theta_n)$ for a total of $25,000$ iterations. The posterior distributions of the parameters are given in Figures 4a-4d.  In Table 1, we give the sample posterior mean together with the $2.5\%$ and $97.5\%$ quantiles of the posterior distribution. Overall, these results are consistent with the maximum likelihood estimates obtained by \cite{robinsetal07} using MCMC-MLE. The main difference appears in $\theta_4$ which we find here not to be significant. As a by-product, the sampler gives an estimate of the variance-covariance matrix of the posterior distribution $\pi$:

\begin{equation}\Sigma_\pi=\left[\begin{tabular}{cccc}1.67 & -0.41 & 0.27 & -0.07\\-0.41 & 1.83 & -0.47 & -0.02 \\0.27 & -0.47 & 1.78 & -0.03\\-0.07 & -0.02 & -0.03 & 1.65\end{tabular}\right].\end{equation}

\vspace{0.5cm}

\small
\begin{table}[h]\label{table1}
\begin{center}
\begin{tabular}{lcc}
\hline
Parameters & Post. mean & Post. quantiles\\
\hline
$\theta_1$ & $-2.14$ & $(-3.32, -0.81)$\\
$\theta_2$ & $0.94$ & $(-0.43, 2.49)$\\
$\theta_3$ & $-1.06$ & $(-2.72, 0.04)$\\
$\theta_4$ & $0.09$ & $(-1.39, 1.07)$\\
\hline
\end{tabular}
\caption{\footnotesize Summary of the posterior distribution of the parameters. Posterior means, $2.5\%$ and $97.5\%$ quantiles}
\end{center}
\end{table}
\normalsize

\begin{center}
\scalebox{0.7}{\includegraphics{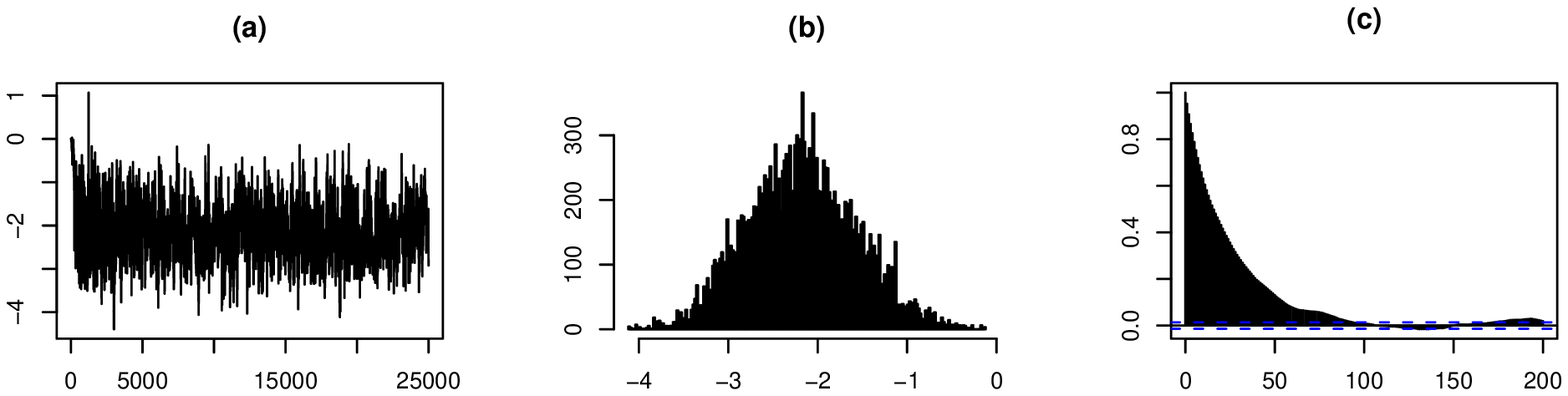}}\\
\noindent\underline{Figure 4a}: The adaptive MCMC output from (\ref{pisn}). (a)-(c): Plots for $\{\theta_{1}\}$. Based on $25,000$ iterations. 
\end{center} 

\begin{center}
\scalebox{0.7}{\includegraphics{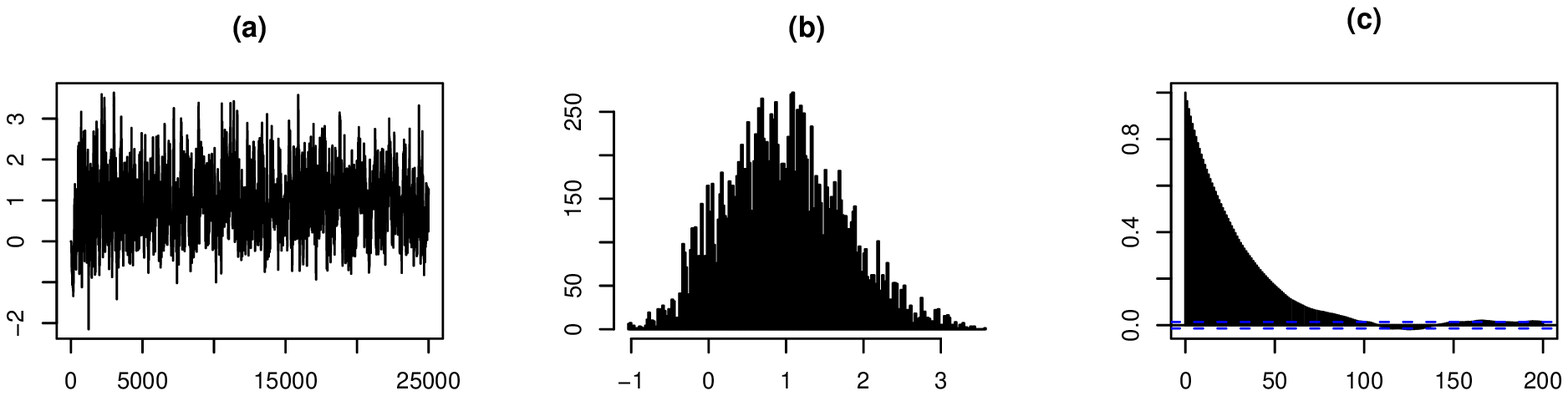}}\\
\noindent\underline{Figure 4b}: The adaptive MCMC output from (\ref{pisn}). (a)-(c): Plots for $\{\theta_{2}\}$. Based on $25,000$ iterations.
\end{center} 
\begin{center}
\scalebox{0.7}{\includegraphics{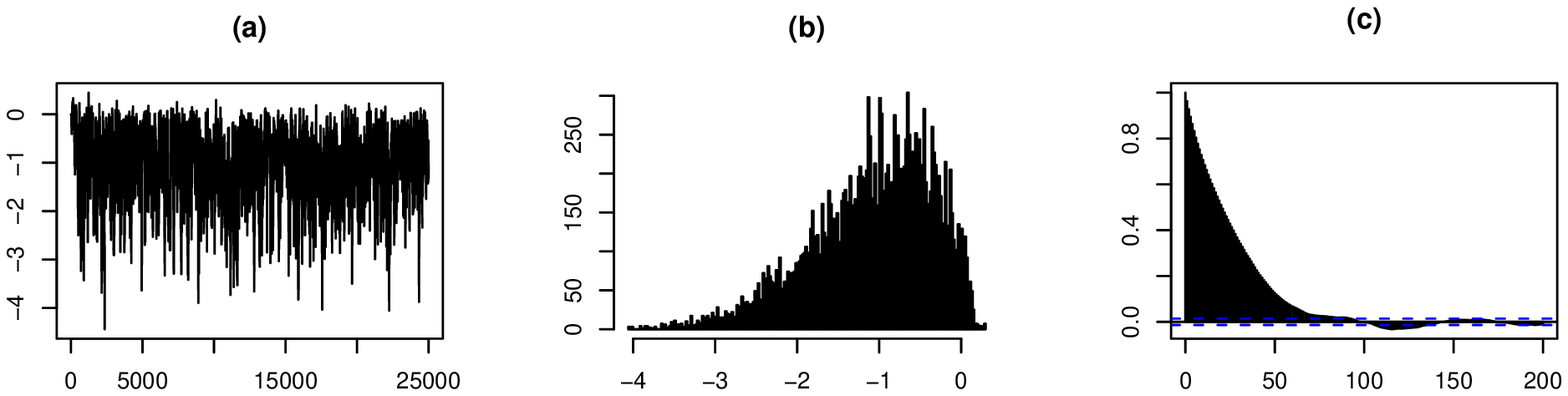}}\\
\noindent\underline{Figure 4c}: The adaptive MCMC output from (\ref{pisn}). (a)-(c): Plots for $\{\theta_{3}\}$. Based on $25,000$ iterations.
\end{center} 
\begin{center}
\scalebox{0.7}{\includegraphics{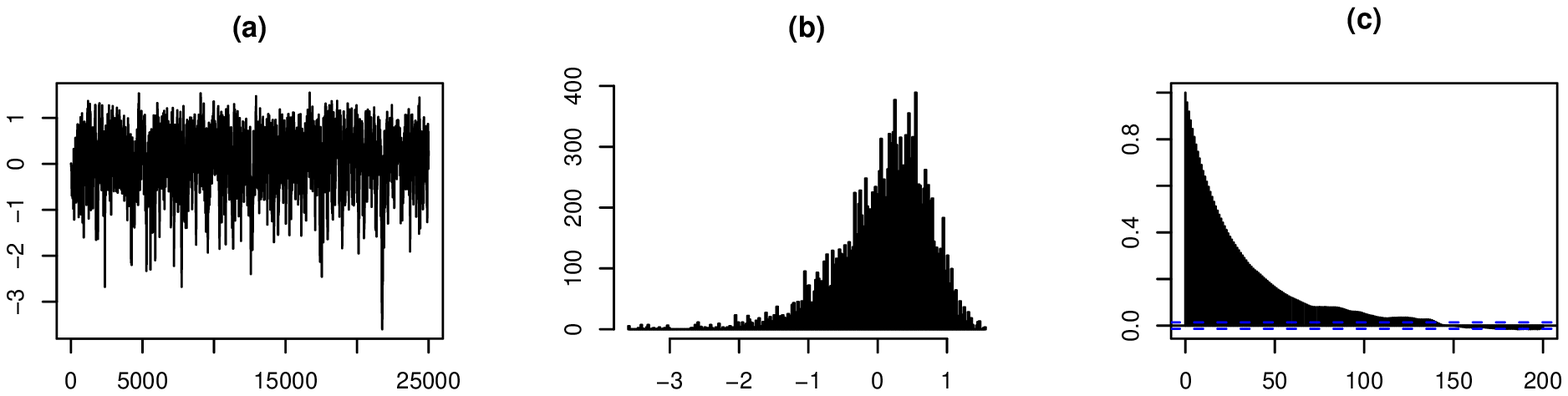}}\\
\noindent\underline{Figure 4d}: The adaptive MCMC output from (\ref{pisn}). (a)-(c): Plots for $\{\theta_{4}\}$. Based on $25,000$ iterations.
\end{center}

\section{Conclusion}\label{conclusion}
Sampling from posterior distributions with intractable normalizing constants is a difficult computational problem. Thus far, all methods proposed in the literature but one entail approximations that do not vanish asymptotically. And the only exception (\cite{molleretal06}) requires exact sampling in the data space, which is only possible for very specific cases. In this work, we propose an approach that both is more general than \cite{molleretal06} and satisfies a strong law of large numbers with limiting distribution equal to the target distribution. The few applications we have presented here suggest that the method is promising. It remains to be determined how the method will scale with the dimensionality and with the size of the problems, although in this respect, adaptations of the method involving annealing/tempering schemes are easy to imagine, which would allow large problems to be analysed properly.

\section*{Acknowledgements}

The research of the third author had been partly supported by the Agence Nationale de la Recherche (ANR, 212,
rue de Bercy 75012 Paris) through the 2006-2008 project {\sf Adap'MC}.

\section{Proof of Theorem \ref{thm1}}\label{proofs}
\begin{proof}
Throughtout the proof, $C$ will denote a finite constant but whose actual value can change from one equation to the next. The convergence of the Wang-Landau algorithm has been studied in \cite{atchadeetliu04}. It is shown in this work that under the condition of Theorem \ref{thm1}, $\min_{i}\sum_{k=1}^\infty\textbf{1}_{\{i\}}(I_k)=\infty$ and more importantly, $e^{c_n(i)}/\sum_{j=1}^de^{c_n(j)}$ converges almost surely to a $Z(\theta^{(i)})$ (up to a multiplicative constant).

Define \[\omega_n(i)=\frac{e^{c_n(i)}}{\sum_{j=1}^de^{c_n(j)}},\]
\[v_{n,i}(\theta)=\frac{\sum_{k=1}^{n}e^{E(X_k,\theta)-E(X_k,\theta^{(i)})}\textbf{1}_{i}(I_k)}{\sum_{k=1}^{n}\textbf{1}_{i}(I_k)}\] 
and 
\begin{equation}\label{zntilde}\tilde Z_n(\theta):=\frac{Z_n(\theta)}{\sum_{j=1}^de^{c_n(j)}}=\sum_{i=1}^d\omega_n(i)v_{n,i}(\theta).\end{equation}
Instead of $Z_n$, we work with $\tilde Z_n$. This is equivalent because $\sum_{j=1}^de^{c_n(j)}$ does not depend on $\theta$ and $Z_n$ always appears in $Q_{Z_n}$ as a ratio. We have:
\begin{equation}\label{zn1} \inf_{\theta\in\Theta}\tilde Z_n(\theta)\geq e^{m-M}.\end{equation}
\begin{equation}\label{zn2} \inf_{\theta,\theta'\in\Theta}\left(\frac{\tilde Z_n(\theta)}{\tilde Z_n(\theta')}\right)\geq e^{2(m-M)}.\end{equation}
Combining (\ref{zn2}) and (\ref{a2}) and part 2 of Remark \ref{rema2}, we deduce that there exists $\eps_0>0$ such that for all $n,j\geq 0$
\begin{equation}\label{unifgeoergo}\sup_{\abs{h}\leq 1}\abs{Q_{Z_n}^jh(\theta)-\pi_{Z_n}(h)}\leq 2(1-\eps_0)^j,\;\;\Pr-a.s.\end{equation}
We introduce the notation $\bar Q_n=Q_{\tilde Z_n}-\pi_{\tilde Z_n}$. It follows from (\ref{unifgeoergo}) that for any $n\geq 1$ the following function $g_n$ is well defined:
\[g_n(\theta)=\sum_{j=1}^\infty\bar Q^j_nh(\theta).\]
Moreover $\abs{g_n(\theta)}\leq 2/\eps_0$ for all $\theta\in\Theta$. $g_n$ satisfies Poisson's equation for $\bar Q_n$ and $h-\pi_{\tilde Z_n}(h)$:
\begin{equation}\label{peq}g_n(\theta)-\bar Q_ng_n(\theta)=h-\pi_{Z_n}(h).\end{equation}
Using this we can rewrite $\sum_{k=1}^nh(\theta_k)-\pi_{Z_k}(h)$ as:
\begin{eqnarray}\frac{1}{n}\sum_{k=1}^n\left(h(\theta_k)-\pi_{Z_k}(h)\right)&=&\frac{1}{n}\sum_{k=1}^n\left(g_k(\theta_k)-\bar Q_{k}g_k(\theta_{k-1})\right)+\frac{1}{n}\sum_{k=1}^n\left(\bar Q_kg_k(\theta_{k-1})-\bar Q_{k-1}g_{k-1}(\theta_{k-1})\right)\nonumber\\
&&+\frac{1}{n}\left(\bar Q_0g_0(\theta_{0})-\bar Q_{n}g_{n}(\theta_{n})\right).\end{eqnarray}

Since $\sup_{\theta\in\Theta}\abs{g_n(\theta)}\leq 2/\eps_0$, a similar bound hold for $\bar Q_ng_n$ and we conclude that $\frac{1}{n}\left(\bar Q_0g_0(\theta_{0})-\bar Q_{n}g_{n}(\theta_{n})\right)$ actually converges almost surely to $0$ as $n\to\infty$.
Writing
$D_k=g_k(\theta_k)-\bar Q_{k}g_k(\theta_{k-1})$, it is easily seen that $\{D_k,\F_k\}$ is a martingale difference with bounded increment and we deduce from martingales theory that $\frac{1}{n}\sum_{k=1}^n\left(g_k(\theta_k)-\bar Q_{k}g_k(\theta_{k-1})\right)$ converges almost surely to $0$ as $n\to\infty$.

Since $Q_Z$ is a Metropolis kernel and using the fact that $\abs{\min(1,ax)-\min(1,ay)}\leq a\abs{x-y}$ for all $a,x,y\geq 0$ we deduce that for any function $h:\;\Theta\to\Real$ such that $\abs{h}\leq 1$, 
\begin{eqnarray}\label{stab}\abs{(Q_{\tilde Z_n}-Q_{\tilde Z_{n-1}})h(\theta)}&\leq&\int\abs{\frac{\tilde Z_{n}(\theta)}{\tilde Z_n(\theta')}-\frac{\tilde Z_{n-1}(\theta)}{\tilde Z_{n-1}(\theta')}}e^{E(x_0,\theta')-E(x_0,\theta)}p(\theta,\theta')\abs{h(\theta')-h(\theta)}d\theta'\nonumber\\
 &\leq&2e^{M-m}\sup_{\theta,\theta'\in\Theta}\abs{\frac{\tilde Z_{n}(\theta)}{\tilde Z_n(\theta')}-\frac{\tilde Z_{n-1}(\theta)}{\tilde Z_{n-1}(\theta')}}\nonumber\\
&\leq& C\abs{\tilde Z_{n}(\theta)-\tilde Z_{n-1}(\theta)},\;\;\mbox{ using } (\ref{zn1}-\ref{zn2}),\end{eqnarray}
for some finite constant. Combining (\ref{unifgeoergo} and \ref{stab}) we have the following well-known consequence: there exists $C<\infty$ such that for all $n\geq 1$:
\begin{equation}\label{stabpi}\sup_{\abs{h}\leq 1}\abs{\pi_{\tilde Z_n}(h)-\pi_{\tilde Z_{n-1}}(h)}\leq C\sup_{\theta\in\Theta}\abs{\tilde Z_n(\theta)-\tilde Z_{n-1}(\theta)}.\end{equation}
The stability of Poisson's equation for geometrically ergodic transition kernels is well known (see e.g. \cite{andrieuetal06, atchadeetliu04}). Combining (\ref{unifgeoergo}), (\ref{stab}) and (\ref{stabpi}), we can find a finite constant $C$ such that for all $k\geq 1$:
\begin{equation}\abs{\left(\bar Q_kg_k(\theta_{k-1})-\bar Q_{k-1}g_{k-1}(\theta_{k-1})\right)}\leq C\sup_{\theta,\theta'\in\Theta}\abs{\tilde Z_k(\theta)-\tilde Z_{k-1}(\theta)}.\end{equation}

Given the expression of $\tilde Z_n(\theta)$ in (\ref{zntilde}) it is not very hard to show there exists $C<\infty$ such that:
\begin{equation}\label{zn3}\sup_{\theta\in\Theta}\abs{\tilde Z_k(\theta)-\tilde Z_{k-1}(\theta)}\leq C\left(d\gamma_k+\frac{1}{\min_i\sum_{l=1}^k\textbf{1}_{\{i\}}(I_l)}\right)\to 0,\end{equation}
as $k\to\infty$ as discussed above. It follows indeed that $\frac{1}{n}\sum_{k=1}^n\left(h(\theta_k)-\pi_{Z_k}(h)\right)$ converges a.s. to $0$.

Given that $\tilde Z_n(\theta)\to CZ(\theta)$ almost surely for some finite constant $C$,
\begin{equation*}\pi_{Z_n}(h)=\frac{\int\frac{e^{E(\theta,x_0)}}{Z_n(\theta)}h(\theta)d\theta}{\int\frac{e^{E(\theta,x_0)}}{Z_n(\theta)}d\theta}\To\frac{\int\frac{e^{E(\theta,x_0)}}{Z(\theta)}h(\theta)d\theta}{\int\frac{e^{E(\theta,x_0)}}{Z(\theta)}d\theta}=\pi(h),\end{equation*}
as $n\to\infty$ by Lebesgue's dominated convergence.
\end{proof}

\bibliographystyle{ims}
\bibliography{ALR08}

\begin{thebibliography}{18}
\expandafter\ifx\csname natexlab\endcsname\relax\def\natexlab#1{#1}\fi
\expandafter\ifx\csname url\endcsname\relax
  \def\url#1{\texttt{#1}}\fi
\expandafter\ifx\csname urlprefix\endcsname\relax\def\urlprefix{URL }\fi
\providecommand{\eprint}[2][]{\url{#2}}

\bibitem[{Andrieu and Moulines(2006)}]{andrieuetal06}
\textsc{Andrieu, C.} and \textsc{Moulines, {\'E}.} (2006).
\newblock On the ergodicity properties of some adaptive {MCMC} algorithms.
\newblock \textit{Ann. Appl. Probab.}, \textbf{16} 1462--1505.

\bibitem[{Atchade(2006)}]{atchade05}
\textsc{Atchade, Y.~F.} (2006).
\newblock An adaptive version for the {M}etropolis adjusted {L}angevin
  algorithm with a truncated drift.
\newblock \textit{Methodol Comput Appl Probab}, \textbf{8} 235--254.

\bibitem[{Atchade and Liu(2004)}]{atchadeetliu04}
\textsc{Atchade, Y.~F.} and \textsc{Liu, J.~S.} (2004).
\newblock The {W}ang-{L}andau algorithm for {M}onte {C}arlo computation in
  general state spaces.
\newblock \textit{Technical Report}.

\bibitem[{Besag(1974)}]{besag74}
\textsc{Besag, J.} (1974).
\newblock Spatial interaction and the statistical analysis of lattice systems.
\newblock \textit{J. Roy. Statist. Soc. Ser. B}, \textbf{36} 192--236.
\newblock With discussion by D. R. Cox, A. G. Hawkes, P. Clifford, P. Whittle,
  K. Ord, R. Mead, J. M. Hammersley, and M. S. Bartlett and with a reply by the
  author.

\bibitem[{Cucala et~al.(2008)Cucala, Marin, Robert and
  Titterington}]{marinetal07}
\textsc{Cucala, L.}, \textsc{Marin, J.-M.}, \textsc{Robert, C.} and
  \textsc{Titterington, D.} (2008).
\newblock A {B}ayesian reassessment of nearest-neighbour classification.
\newblock Tech. rep., CEREMADE, Universit{\'e} Paris Dauphine.
\newblock \eprint{arXiv:0802.1357}.

\bibitem[{Gelman and Meng(1998)}]{gelmanetmeng98}
\textsc{Gelman, A.} and \textsc{Meng, X.-L.} (1998).
\newblock Simulating normalizing constants: from importance sampling to bridge
  sampling to path sampling.
\newblock \textit{Statist. Sci.}, \textbf{13} 163--185.

\bibitem[{Geyer(1994)}]{geyer94}
\textsc{Geyer, C.~J.} (1994).
\newblock On the convergence of {M}onte {C}arlo maximum likelihood
  calculations.
\newblock \textit{J. Roy. Statist. Soc. Ser. B}, \textbf{56} 261--274.

\bibitem[{Geyer and Thompson(1992)}]{geyeretthompson92}
\textsc{Geyer, C.~J.} and \textsc{Thompson, E.~A.} (1992).
\newblock Constrained {M}onte {C}arlo maximum likelihood for dependent data.
\newblock \textit{J. Roy. Statist. Soc. Ser. B}, \textbf{54} 657--699.
\newblock With discussion and a reply by the authors.

\bibitem[{Hurn et~al.(2003)Hurn, Husby and Rue}]{hurnetal03}
\textsc{Hurn, M.}, \textsc{Husby, O.} and \textsc{Rue, H.} (2003).
\newblock A tutorial on image analysis.
\newblock \textit{Lecture notes in Statistics}, \textbf{173} 87--141.

\bibitem[{Ibanez and Simo(2003)}]{ibanezetsimo03}
\textsc{Ibanez, M.~V.} and \textsc{Simo, A.} (2003).
\newblock Parameter estimation in {M}arkov random field image modeling with
  imperfect observations. a comparative study.
\newblock \textit{Pattern recognition letters}, \textbf{24} 2377--2389.

\bibitem[{Kleinman et~al.(2006)Kleinman, Rodrigue, Bonnard and
  Philippe}]{lartillotetal06}
\textsc{Kleinman, A.}, \textsc{Rodrigue, N.}, \textsc{Bonnard, C.} and
  \textsc{Philippe, H.} (2006).
\newblock A maximum likelihood framework for protein design.
\newblock \textit{BMC Bioinformatics}, \textbf{7}.

\bibitem[{M{\o}ller et~al.(2006)M{\o}ller, Pettitt, Reeves and
  Berthelsen}]{molleretal06}
\textsc{M{\o}ller, J.}, \textsc{Pettitt, A.~N.}, \textsc{Reeves, R.} and
  \textsc{Berthelsen, K.~K.} (2006).
\newblock An efficient {M}arkov chain {M}onte {C}arlo method for distributions
  with intractable normalising constants.
\newblock \textit{Biometrika}, \textbf{93} 451--458.

\bibitem[{M{\o}ller and Waagepetersen(2004)}]{molleretwaag03}
\textsc{M{\o}ller, J.} and \textsc{Waagepetersen, R.~P.} (2004).
\newblock \textit{Statistical inference and simulation for spatial point
  processes}, vol. 100 of \textit{Monographs on Statistics and Applied
  Probability}.
\newblock Chapman \& Hall/CRC, Boca Raton, FL.

\bibitem[{Murray et~al.(2006)Murray, Ghahramani and MacKay}]{murrayetal06}
\textsc{Murray, I.}, \textsc{Ghahramani, Z.} and \textsc{MacKay, D.} (2006).
\newblock {MCMC} for doubly-intractable distributions.
\newblock \textit{Proceedings of the 22nd Annual Conference on Uncertainty in
  Artificial Intelligence (UAI)}.

\bibitem[{Plagnol and Tavar{\'e}(2004)}]{plagnolettavare04}
\textsc{Plagnol, V.} and \textsc{Tavar{\'e}, S.} (2004).
\newblock Approximate {B}ayesian computation and {MCMC}.
\newblock In \textit{{M}onte {C}arlo and quasi-{M}onte {C}arlo methods 2002}.
  Springer, Berlin, 99--113.

\bibitem[{Robins et~al.(2007)Robins, P., Kalish and Lusher}]{robinsetal07}
\textsc{Robins, G.}, \textsc{P., P.}, \textsc{Kalish, Y.} and \textsc{Lusher,
  D.} (2007).
\newblock An introduction to exponential random graph models for social
  networks.
\newblock \textit{Social Networks}, \textbf{29} 173--191.

\bibitem[{Wang and Landau(2001)}]{wangetlandau01}
\textsc{Wang, F.} and \textsc{Landau, D.~P.} (2001).
\newblock Efficient, multiple-range random walk algorithm to calculate the
  density of states.
\newblock \textit{Physical Review Letters}, \textbf{86} 2050--2053.

\bibitem[{Younes(1988)}]{younes88}
\textsc{Younes, L.} (1988).
\newblock Estimation and annealing for {G}ibbsian fields.
\newblock \textit{Annales de l'Institut Henri Poincar\'e. Probabilit\'e et
  Statistiques}, \textbf{24} 269--294.

\end{thebibliography}

\end{document}